\documentclass[aps,prd,preprint,nofootinbib]{revtex4-1}
\pdfoutput=1 
\usepackage{graphicx}
\usepackage{color}
\usepackage{rotating}
\usepackage{bigints}
\usepackage{enumerate}

\newcommand{\beq}{\begin{equation}}
\newcommand{\eeq}{\end{equation}}
\newcommand{\bea}{\begin{eqnarray}}
\newcommand{\eea}{\end{eqnarray}}

\begin{document}
\title{Future Physics Perspectives on the Equation of State from Heavy Ion Collisions to Neutron Stars}

\author{V.~Dexheimer}
\email{vdexheim@kent.edu}
\affiliation{Department of Physics, Kent State University, Kent, OH 44243 USA}

\author{J.~Noronha}
\email{jn0508@illinois.edu}
\affiliation{Illinois  Center  for  Advanced  Studies  of  the  Universe, Department of Physics, 
University of Illinois at Urbana-Champaign, Urbana, IL 61801, USA}

\author{J.~Noronha-Hostler}
\email{jnorhos@illinois.edu}
\affiliation{Illinois  Center  for  Advanced  Studies  of  the  Universe, Department of Physics, 
University of Illinois at Urbana-Champaign, Urbana, IL 61801, USA}

\author{C.~Ratti}
\email{cratti@central.uh.edu}
\affiliation{Department of Physics, University of Houston, Houston, TX, USA  77204}

\author{N.~Yunes}
\email{nyunes@illinois.edu}
\affiliation{Illinois  Center  for  Advanced  Studies  of  the  Universe, Department of Physics, 
University of Illinois at Urbana-Champaign, Urbana, IL 61801, USA}

\date{\today}

\begin{abstract}
With the computational power and algorithmic improvements available today, the ongoing STAR/RHIC and HADES/GSI experiments, the future FAIR and NICA facilities becoming operational, and the new precise measurements from NICER and LIGO/VIRGO, the high-energy nuclear physics and astrophysics communities are in the unique position to set very stringent constraints on the equation of state of strongly interacting matter. We review the state-of-the-art of different approaches used in the description of hot and ultradense baryonic matter in and out of equilibrium, and discuss the regions in the phase diagram where heavy-ion collisions and neutron star mergers can overlap. Future perspectives are discussed to help define a comprehensive, multi-disciplinary strategy to map out the phase diagram of strongly interacting matter from heavy ion collisions to neutron stars.
\end{abstract}

\maketitle
\section{Introduction}

Relativistic heavy ion collisions, such as those taking place at the Relativistic Heavy Ion Collider (RHIC) or at the Large Hadron Collider (LHC), routinely create a new phase of matter called the quark-gluon plasma (QGP). The QGP is formed only at extremely large temperatures and particle/antiparticle densities and it permeated the early Universe just a few microseconds after the Big Bang. The phase transition from hadronic matter to the QGP is a broad crossover when matter and antimatter are present in approximately the same amount \cite{Aoki:2006we}, i.e. at zero baryon chemical potential $\mu_B=0$, but the transition is expected to become first-order at high baryonic densities \cite{Stephanov:2007fk,Nahrgang:2016ayr,Ratti:2018ksb,Bzdak:2019pkr}. This implies the existence of a high-temperature $T$ critical point (CP) \cite{Stephanov:1998dy} on the phase diagram of quantum chromodynamics (QCD). In the next couple of years, questions concerning the potential existence and location of the QCD critical point may finally be answered. The STAR experiment at RHIC is running a second beam energy scan (BESII) until 2022 to find out whether a hot and dense system of quarks and gluons displays critical phenomena when doped with more quarks than antiquarks (finite net density). STAR has also developed a fixed target program to further extend their reach to larger densities \cite{STARnote,Cebra:2014sxa}   and HADES at GSI \cite{Galatyuk:2014vha} covers even larger densities with the potential of a beam energy  scan as well. The search for this putative critical point can also shed new light on whether quark matter exists in dense stellar objects, given that the conditions achieved in low energy heavy ion collisions \cite{Adamczewski-Musch:2019byl} can overlap with those in neutron star mergers \cite{Most:2018eaw}. 

\begin{figure*}
    \centering
    \includegraphics[width=0.6\textwidth]{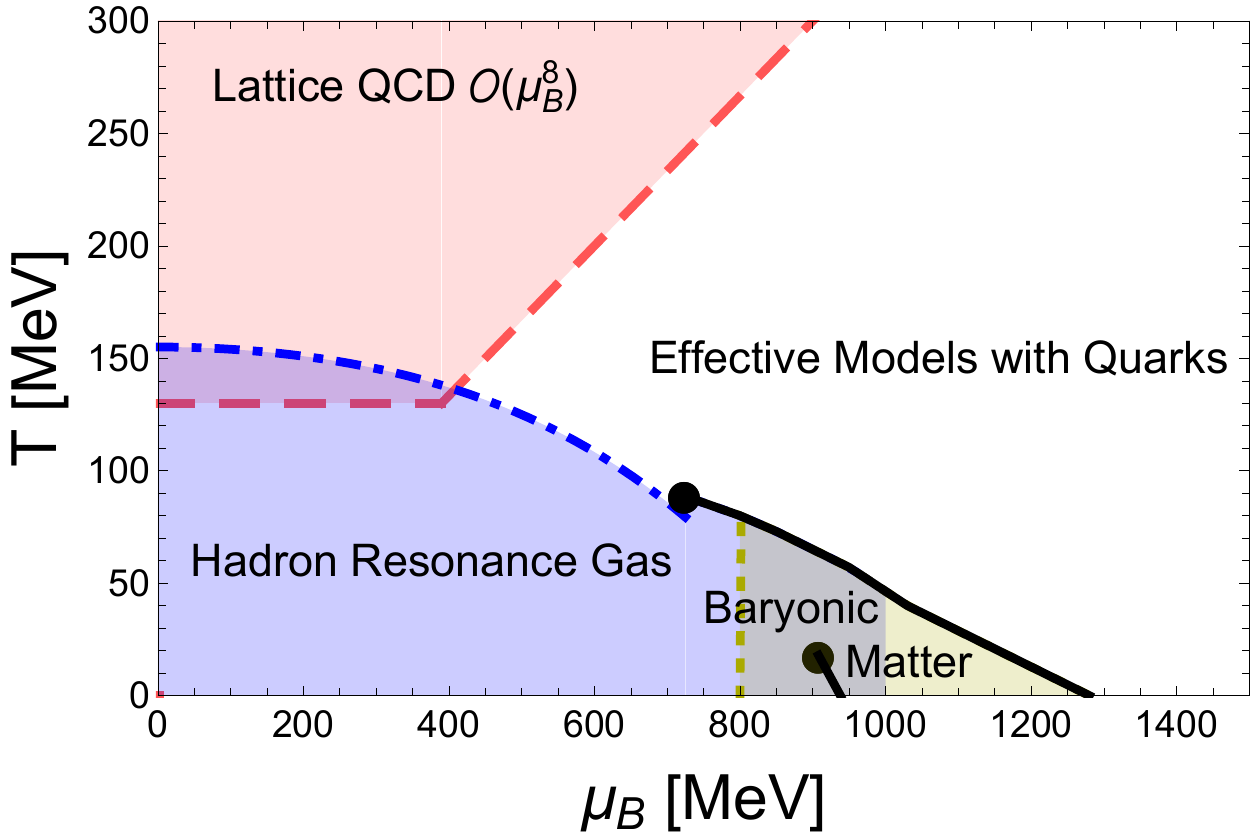}
    \caption{Schematic representation of the QCD phase diagram. Thick lines mark the liquid-gas and quark deconfinement phase transitions. The black dots mark critical points with predictions from \cite{Critelli:2017oub} and \cite{Elliott:2012nr}.}
    \label{Fig:QCDphasediagram}
\end{figure*}

The information provided by low-energy heavy-ion collisions, which will continue after RHIC with the FAIR (GSI) \cite{Friese:2006dj,Tahir:2005zz,Lutz:2009ff,Durante:2019hzd} and NICA (Dubna) \cite{Kekelidze:2017tgp,Kekelidze:2016wkp} programs, can be complemented for the first time by very precise astrophysical observations that allow one to constrain the notoriously difficult high-density/low-temperature corner of the QCD phase diagram (see Fig.\ \ref{Fig:QCDphasediagram}). In fact, the unprecedented accuracy of recent astrophysical experiments, such as NASA's Neutron Star Interior Composition ExploreR (NICER) and NSF's Laser Interferometer Gravitational-Wave Observatory (LIGO), provides a new possibility to test and rule out hypotheses for the neutron star inner core composition, thus constraining the high-energy QCD equation of state (EoS) and its phase diagram. From a theoretical point of view, since QCD cannot be directly solved at high (net) baryon densities because of the sign problem \cite{Troyer:2004ge} (for a review, see \cite{Ratti:2018ksb}), the existence and location of the critical point and the high-density phases of matter described by their EoSs are at the moment not well constrained. These EoSs are not directly comparable to heavy-ion experimental data, but serve as a vital input in event-by-event relativistic viscous hydrodynamic simulations of heavy ion collisions \cite{Alba:2017hhe}. Furthermore, the inclusion of viscous effects in hydrodynamic simulations of the QGP is required to describe heavy ion data \cite{Heinz:2013th}, which in turn provide key information about the transport properties (such as shear and bulk viscosities) of the QGP \cite{Bernhard:2019bmu}. In fact, most theoretical studies indicate that viscous and diffusive effects are even more relevant at large densities \cite{Monnai:2016kud,Auvinen:2017fjw,Fotakis:2019nbq,Dore:2020jye}, as we will discuss in more detail in Sec.\ \ref{sec:dyn}.

Moving towards the high baryon density regime, the EoS of strongly interacting matter dictates stellar masses and radii as determined by the Tolman-Oppenheimer-Volkoff equations \cite{Tolman:1939jz,Oppenheimer:1939ne}. Other stellar observables such as the spindown of fast pulsars \cite{Alford:2013pma}, the cooling rate of a large set of stable neutron stars \cite{Grigorian:2004jq,Page:2005fq,Negreiros:2018cho,Tolos:2016hhl}, the compactness and ellipticity of rotating neutron stars with hot spots on their surface, and tidal deformability measurements that determine the star's quadrupole moment in response to a strong tidal field during the inspiral phase of a neutron star merger \cite{Yagi:2013bca}, also generate important constraints for theoretical calculations \cite{Oertel:2016bki}. In neutron star mergers, knowledge about the EoS at high baryon density and finite temperatures is needed when performing general-relativistic hydrodynamic simulations \cite{Most:2018eaw}. The inclusion of viscous effects in the latter may also be important to determine the evolution of the hypermassive remnant formed after the merger, as stressed in  \cite{Duez:2004nf,Shibata:2017jyf,Shibata:2017xht,Alford:2017rxf,Radice:2018ghv}.

The authors believe that this is the right time to bring together the scientific heavy-ion and neutron-star communities to discuss common goals involving the EoS of strongly interacting matter and define a strategy to achieve them. For this reason, they organized a 3-day virtual workshop in August 2020, supported by NSF, which brought together lattice and perturbative QCD (pQCD) theorists, heavy-ion phenomenologists and experimentalists, nuclear astrophysicists and gravitational wave physicists to discuss the state-of-the-art in their fields and define a path forward to address and solve the most urgent overlapping issues of these communities. Given the diversity of the participants' background, long and lively panel discussion sessions took place in the end of each day of the workshop, during which a moderator and panelists listed the main open questions in each community and discussed ways of addressing them with the other participants. In this document, we review the main topics covered in the workshop and some of the problems and common goals that the communities identified as needing to be urgently addressed.

\section{Equation of state at low-to-moderate densities: status and perspectives}

The EoS of QCD at zero net-baryonic density has been available for a few years in the continuum limit and for physical values of the quark masses, for a system of $N_f=2+1$ \cite{Borsanyi:2010cj,Borsanyi:2013bia,Bazavov:2014pvz} and $N_f=2+1+1$ \cite{Borsanyi:2016ksw} quark flavors. These results agree with the thermodynamics generated by a gas of non-interacting hadrons and resonances at low temperatures \cite{Borsanyi:2012cr,Borsanyi:2014ewa,Alba:2017mqu}, and with a gas of weakly interacting quarks and gluons at high temperatures \cite{Laine:2006cp,Andersen:2011sf,Haque:2014rua}.
In the field of heavy-ion collisions, these results are crucial to describe the fireball created in the collision, as the EoS is one of the key inputs in a hydrodynamic description of the system \cite{Romatschke:2017ejr}. Indeed, an important independent validation of lattice QCD results was obtained in \cite{Pratt:2015zsa}, where state-of-the-art statistical techniques have been applied to the combined analysis of a large number of experimental observables, while changing the input parameters in a controlled way. It was found that the posterior distribution over possible parametrizations of the EoS was consistent with current lattice QCD calculations.

First principles lattice QCD results of the EoS at nonzero baryon chemical potential are currently limited due to the sign problem, a fundamental technical obstacle of exponential complexity present in path integral approaches solved via importance sampling \cite{Troyer:2004ge}. The most common methods to extend the lattice QCD results beyond $\mu_B=0$ are the Taylor expansion of the thermodynamic observables around $\mu_B=0$ \cite{Allton:2002zi,Allton:2005gk,Gavai:2008zr,Basak:2009uv,Kaczmarek:2011zz} and the analytical continuation of simulations performed at imaginary chemical potential \cite{deForcrand:2002hgr,DElia:2002tig,Wu:2006su,DElia:2007bkz,Conradi:2007be,deForcrand:2008vr,DElia:2009pdy,Moscicki:2009id}. Exploratory studies in simplified fermionic models where the sign problem could be solved using other techniques, such as Lefschetz thimbles, can be found for instance in \cite{Alexandru:2015xva,Alexandru:2015sua}.

In general, one can write the pressure $p$ of QCD as a Taylor series in powers of the baryon chemical potential over temperature, $\mu_B/T$, around $\mu_B=0$:
\begin{eqnarray}
\frac{p(T,\mu_B)}{T^4}&=&\frac{p(T,0)}{T^4} +\sum_{n=1}^\infty\left.\frac{1}{n!}\frac{\mathrm{d^{n}}(p/T^4)}{d(\frac{\mu_B}{T})^{n}}\right |_{\mu_B=0}\left(\frac{\mu_B}{T}\right)^{n}
\nonumber\\
&=&\sum_{n=0}^{\infty}c_{n}(T)\left(\frac{\mu_B}{T}\right)^{n},
\end{eqnarray}
with Taylor coefficients given by
\bea
c_n=\frac{1}{n!}\frac{\partial^n(p/T^4)}{\partial(\mu_B/T)^n},
\eea
which determine the baryon susceptibilities \cite{Ratti:2018ksb}. We note that in previous works other types of expansions have been explored (for instance, the Pade approximation) but it was generally found that the Taylor series worked best for our current data set \cite{Critelli:2017oub}. 

It is important to mention that the QCD phase diagram relevant for heavy ion collisions is, in fact, at least a four-dimensional space in the variables $(T,~\mu_B,~\mu_S,~\mu_Q)$, where $\mu_S$ and $\mu_Q$ stand for the strangeness and electric charge chemical potentials, respectively. 
Therefore, when performing a Taylor expansion in $\mu_B$, one has to make a choice about the values of $\mu_S$ and $\mu_Q$, as well when studying global quantities. Two common choices, useful for heavy ion collision physics, are to consider either $\mu_S=\mu_Q=0$ or a situation in which $\mu_S$ and $\mu_Q$ are functions of $T$ and $\mu_B$, such that they satisfy the following phenomenological relations for the strangeness, charge, and baryon densities
\bea
\langle \rho_S\rangle=0~~~~~~\langle \rho_Q\rangle=0.4\,\langle\rho_B\rangle,
\label{phenoconstraints}
\eea
which reflect the initial conditions in heavy-ion collisions and the fact that strangeness and electric charge are conserved in strong interactions.
Results are currently available for the Taylor coefficients up to sixth-order for both conditions \cite{Gunther:2016vcp,Bazavov:2017dus,Borsanyi:2018grb}, together with estimates for $c_8$ at finite lattice spacing in the case where $\mu_S=\mu_Q=0$ \cite{Borsanyi:2018grb}. We point out that, for dynamical simulations, the entire 4D phase diagram is needed due to local fluctuations of conserved charges (e.g. see \cite{Shen:2017ruz}).

More recently, a full four-dimensional EoS has been reconstructed in \cite{Noronha-Hostler:2019ayj}, based on the Taylor expansion coefficients calculated in \cite{Borsanyi:2018grb} up to fourth-order and defined as
\bea
\frac{p(T,\mu_B,\mu_S,\mu_Q)}{T^4}=\sum_{i,j,k}\frac{1}{i!j!k!}\chi^{BSQ}_{ijk}\left(\frac{\mu_B}{T}\right)^i\left(\frac{\mu_S}{T}\right)^j\left(\frac{\mu_Q}{T}\right)^k,
\label{fullT}
\eea
with baryon, strangeness, and electric charge (BSQ) susceptibilities given by
\bea
\chi^{BSQ}_{ijk}=\frac{\partial^{i+j+k}(p/T^4)}{\partial(\frac{\mu_B}{T})^i\partial(\frac{\mu_Q}{T})^j\partial(\frac{\mu_S}{T})^k}.
\eea
Effects from the inclusion of a partial list of sixth-order coefficients were investigated in \cite{Monnai:2019hkn}.

Additionally, with the currently available Taylor expansion coefficients, a reliable EoS from first principles can be reconstructed up to $\mu_B/T\leq2$ \cite{Gunther:2016vcp}.  It was emphasized in the workshop that any realistic EoS should reproduce the lattice QCD one (or smoothly merge with it) in the density regime where the latter is available. We note that, while we have focused on the EoS here, other lattice QCD quantities should also be used to constrain models, such as partial pressures \cite{Alba:2017mqu}, cross-correlators \cite{Bellwied:2019pxh}, and high-order susceptibilities \cite{Borsanyi:2018grb}.
One of the relevant goals for the lattice QCD community is to extend the range of the reconstructed lattice QCD EoS by generating reliable continuum extrapolated higher order coefficients. This needs to be done both for the  extrapolation in $\mu_B$ only, and for the full four-dimensional EoS presented in Eq.\ (\ref{fullT}). 

Finally, we note that at low temperatures, eventually one must match lattice QCD results to a hadron resonance gas model because lattice QCD results can only be calculated down to approximately $T\sim 100-130$ MeV.  While a non-interacting gas of hadrons and resonances works well at small $\mu_B$ (with the exception of higher-order derivatives \cite{Huovinen:2017ogf}), large density calculations require the inclusion of interactions. These interactions are commonly modeled using a van der Waals description, which can in addition describe the nuclear liquid-gas phase transition \cite{Vovchenko:2015vxa,Vovchenko:2017zpj}. This points to the potential interplay between the high $T$ critical point and the liquid-gas phase transition, as already shown in \cite{Steinheimer:2010ib}.

As mentioned before, one of the most relevant still open questions is whether the phase transition of QCD becomes first-order as the  baryon chemical potential/density increases. 
Experimental efforts to find the location of the critical point/first order phase transition coexistence line include the STAR Fixed-Target program ($\sqrt{s_{NN}}=3-7.7$ GeV) \cite{STARnote,Cebra:2014sxa}, HADES at GSI  ($\sqrt{s_{NN}}=1-3$ GeV) \cite{Galatyuk:2014vha}, FAIR at GSI ($\sqrt{s_{NN}}=4.5-9.3$ GeV) \cite{Friese:2006dj,Tahir:2005zz,Lutz:2009ff,Durante:2019hzd}, and NICA ($\sqrt{s_{NN}}=3-5$ GeV) \cite{Kekelidze:2017tgp,Kekelidze:2016wkp}.
The results of these experiments have consequences for the physics of neutron stars and their mergers \cite{TheLIGOScientific:2017qsa,Hanauske:2019vsz}. First-order phase transitions are signaled by a region  where the speed of sound is equal to zero. If a strong first order phase transition between hadronic matter and quark matter was experienced within a neutron star merger, this would have consequences for the ringdown and the remnants \cite{Most:2018eaw,Bauswein:2018bma,Weih:2019xvw,Gieg:2019yzq,Tsokaros:2019anx,Ecker:2019xrw,Blacker:2020nlq,Pang:2020ilf}. In addition, if a strong first-order line occurs at zero temperature, it is possible that mass  twins could be produced wherein two neutron stars have the same mass but very different radii \cite{Alford:2013aca,Dexheimer:2014pea,Benic:2014jia,Montana:2018bkb,Christian:2017jni}. Mass twins could be detected with NICER and LIGO observations of the radius and tidal deformability of neutron stars, respectively.

%
%

The effect of a potential critical point on heavy-ion collision observables was recently reviewed in  \cite{Bzdak:2019pkr}. One of the most prominent experimental signatures is the expected divergence of higher order baryon number fluctuations in the vicinity of the critical point \cite{Stephanov:2008qz}. While critical slowing down phenomena are expected to reduce this effect \cite{Berdnikov:1999ph}, an increase of the kurtosis as the critical point is approached is still expected. A change in monotonicity for the kurtosis as a function of the collision energy was also suggested as a possible critical point signature \cite{Stephanov:2011pb}, although it was pointed out more recently that this non-monotonic behavior is washed out at temperatures immediately below the phase transition and, therefore, it is not likely to be observed in experiments \cite{Mroczek:2020rpm}.

\begin{figure*}
    \centering
    \includegraphics[width=\textwidth]{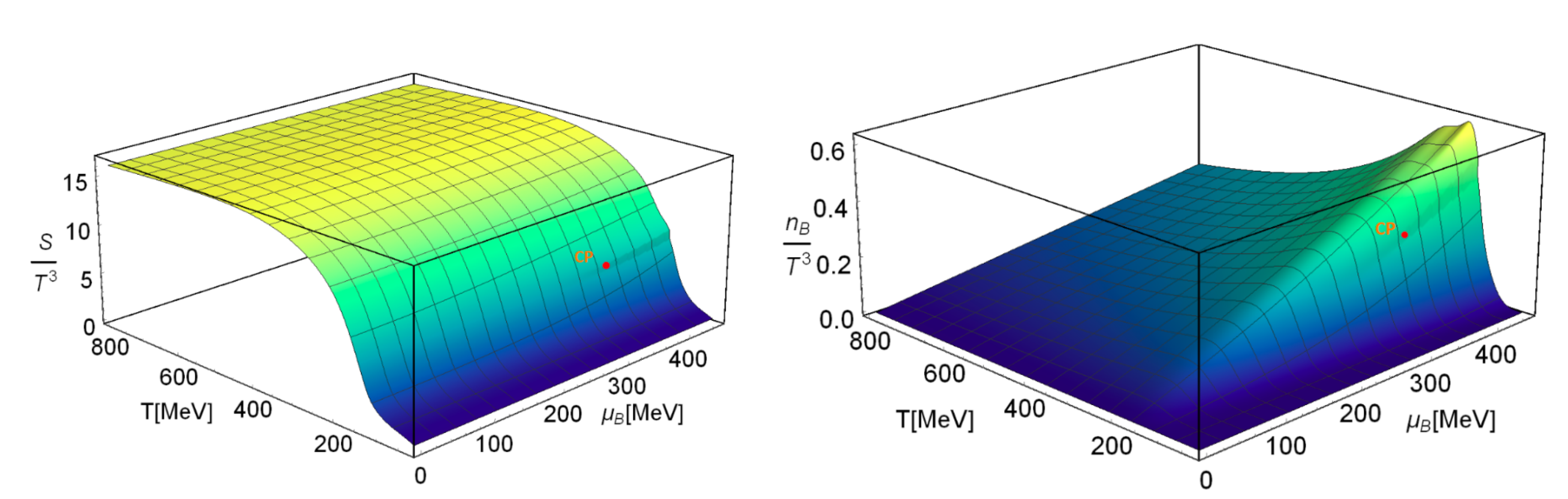}
    \caption{Figure from \cite{Parotto:2018pwx} showing entropy density (left) and baryonic density (right) as functions of the temperature $T$ and chemical potential $\mu_B$. The critical point is marked as a red dot.}
    \label{Fig1}
\end{figure*}

To help answer the question about the existence and location  of the QCD critical point, a family of EoSs was recently constructed in \cite{Parotto:2018pwx}. It reproduces the lattice QCD EoS where it is available and contains a critical point belonging to the universality class of the 3D-Ising model. The uniqueness of this  family of EoSs is that the position and strength of the critical point can be changed and its effect on the data can be explored \cite{Mroczek:2020rpm}. The corresponding entropy density and baryonic density are shown in Fig.\ \ref{Fig1} for one possible choice of parameters. Both quantities are discontinuous for chemical potentials larger than the critical one. Currently, this family of EoSs is available only for the case where $\mu_S=\mu_Q=0$. Extensions to the strangeness-neutral case with fixed charge density, one of the identified priorities for both the heavy-ion and neutron-star communities, are the next obvious steps. This can be achieved using the Taylor series in Eq.\ (\ref{fullT}). 

During the workshop panel discussions, the fate of the critical point in the four-dimensional phase diagram was thoroughly discussed. Given the lack of information from first principle simulations, guidance can be provided by models that reproduce lattice QCD results and contain a critical point such as, for instance, the holographic models in \cite{DeWolfe:2010he,DeWolfe:2011ts}. In particular, the holographic model of Ref.\ \cite{Critelli:2017oub} not only naturally incorporates the nearly perfect fluidity of the QGP \cite{Kovtun:2004de} and allows for the realistic determination of a number of characteristic temperatures \cite{Rougemont:2017tlu} and transport coefficients \cite{Finazzo:2014cna,Rougemont:2015wca,Rougemont:2015ona} as a function of $T$ and $\mu_B$ up to the critical point, but it also successfully predicts the temperature dependence of higher order baryon susceptibilities later computed on the lattice \cite{Borsanyi:2018grb}.   The success of the holographic model \cite{Critelli:2017oub} in comparison to lattice QCD establishes benchmarks that other effective models must pass in order to be consistent with QCD results at finite temperature and density.  

It was pointed out in the workshop discussion that the different high-energy communities explore different regions of the phase diagram, not only in terms of net-baryonic density, but also in terms of the conditions on the other conserved charges (besides baryon number). In particular, while isospin asymmetry is small in heavy ion collisions due to the fact that the initial nuclei contain almost the same amount of protons and neutrons, it is large in neutron stars as a result of the (proton) electron capture that takes place during supernovae. In cold neutron stars, $\mu_Q$ is determined in order to ensure chemical equilibrium and charge neutrality.
On the other hand, while strangeness is a conserved quantity on the timescales of a heavy ion collision, weak decays are significant on astrophysical timescales ($\mu_S=0$) and strangeness is expected to be significant in cold neutron stars. This gives rise to well-defined patterns in the four-dimensional phase diagram, which should be addressed as a first step to obtain a more complete understanding of it.  In fact, different regimes of the phase diagram can experience the transition from quarks to hadrons at different chemical potentials depending on the underlying assumptions \cite{Fu:2018qsk,Rennecke:2019dxt,Aryal:2020ocm}. Nevertheless, we remark that in non-equilibrium situations, such as those involving neutron star merger simulations, a higher dimensional EoS would also be needed. 

Additional questions were raised in the workshop about the influence of multiple conserved charges on the location of the critical point. For instance, in 3 dimensions, does it become a critical line? Does it turn into a critical plane in 4 dimensions?  Or could it be a crossover in certain regions of the phase diagram and a real phase transition in other regions?  While effective models provide some guidance in this regard, the answer currently depends on the underlying assumptions of the models and, therefore, further experimental studies and efforts to extend lattice QCD to large densities are needed. 



\section{Equation of state at high densities and  comparisons with observations}


Cold neutron stars can be well-approximated by an EoS at $T\sim 0$ MeV, covering all scales relevant to nuclear and particle physics, from nuclei at very low densities (crust) to potentially deconfined quarks in the inner core. A review of the different models used to describe the EoS of neutron stars can be found in \cite{Baym:2017whm}. Fundamental questions remain about N-body interactions  between  baryons \cite{Tolos:2007bh,Hebeler:2013nza,Tews:2018kmu,Lim:2018bkq}, the existence of hyperons  (and their interactions) \cite{Weissenborn:2011kb,Weissenborn:2011ut,Lopes:2020rqn,Gerstung:2020ktv}, and the possibility  of deconfinement to quark matter in the core of a neutron star.

At zero temperature, protons and neutrons begin to overlap at densities around 3 times saturation density ($\sim$ 0.5 baryons per fm$^3$), at which point a simple hadronic description of dense matter is not enough. Although the exact position and manner  such transition takes place is still not certain, there are indications from a matching of low energy constraints with pQCD at asymptotically large energies that it is a steep first-order phase transition \cite{Annala:2019puf}. This feature is also hinted by a necessary bump in the dense-matter speed of sound that surpasses the conformal limit and then returns to $c_s^2\rightarrow \frac{1}{3}$ at large densities \cite{Bedaque:2014sqa,Alford:2015dpa,Tews:2018kmu,McLerran:2018hbz,Baym:2019iky,Jakobus:2020nxw}, although this effect could also be caused by a sudden appearance of hyperons (see Fig.~2 of \cite{Stone:2019abq} and \cite{Gulminelli:2013qr}). At sufficiently large densities and low temperatures, quark matter in neutron stars is expected to be a color superconductor, for which many kinds of pairing patterns exist \cite{Alford:2007xm}, including the color-flavor-locked phase  \cite{Alford:1998mk}.

Although the crust of neutron stars is reasonably understood \cite{Chamel:2008ca}, the number of constraints for bulk hadronic matter decreases as the density increases in the core, especially beyond a couple of times saturation density. Up to this point, there are reliable ab-initio calculations (such as CEFT \cite{Hebeler:2013nza,Tews:2018kmu,Lim:2018bkq}), which solve a many-body problem starting from 2- and 3-body interactions, together with laboratory data of saturation properties (see a summary in Fig.~1 of \cite{Li:2013ola}). 
Beyond that point, uncertainties increase, especially due to the appearance of new degrees of freedom such as hyperons and meson condensates \cite{Glendenning:1982nc,Kaplan:1986yq,Baym:1978sz}. In this case, a natural solution is to rely on QCD-inspired relativistic phenomenological models that include both deconfinement and chiral symmetry restoration \cite{Dexheimer:2009hi,Steinheimer:2011ea,Turko:2014jta,Marczenko:2020jma}.

For the quark phase, this is especially problematic, as the large uncertainty in modelling and parameters allow the EoS to be very similar to the hadronic one, which has been referred to as the ``masquerade effect” \cite{Alford:2004pf,Wei:2018mxy}. Of course, at this point one could once more invoke comparisons with pQCD results in the relevant regime \cite{Andersen:2002jz,Fraga:2013qra,Annala:2019puf}, but one must keep in mind that this regime is not achieved inside neutron stars.  pQCD is expected to be applicable at approximately an order of magnitude larger densities compared to the cores of neutron stars, which may only reach 4-10 times nuclear saturation density.  

This is exactly where multi-messenger astrophysical observations can provide guidance to nuclear physics. Measurements of neutron star masses larger than 2 solar masses, together with small radii, and small tidal deformability, imply that the cold matter EoS is soft at intermediate densities but stiff at large densities, which allows us to extract information about nuclear interactions \cite{Dexheimer:2018dhb,Hornick:2018kfi,Dexheimer:2020rlp,Otto:2020hoz,Kubis:2020ysv,Ferreira:2020evu}.  In particular, a neutron star with mass larger than 2.5 solar masses (for instance, the secondary compact object from GW190814 \cite{Abbott:2020khf}) would imply that a sharp rise in $c_
s^2$ occurs between 2-3 times nuclear saturation density \cite{Tan:2020ics}, which may be attributed to exotic degrees of freedom in the core of a neutron star. However, we point out that many alternatives have been suggested for the secondary from GW190814: it may be a black hole or a fast spinning neutron star \cite{Most:2020bba,Dexheimer:2020rlp,Zhang:2020zsc, Tsokaros:2020hli}, or even a primordial black hole \cite{Vattis:2020iuz}. Nevertheless, non-monotonic behavior of $c_
s^2$ has been found in several different works \cite{Dexheimer:2007mt,Bedaque:2014sqa,Alford:2015dpa,Dutra:2015hxa,Ranea-Sandoval:2015ldr,Dexheimer:2017nse,Tews:2018kmu,Tews:2018iwm,McLerran:2018hbz,Jakobus:2020nxw,Zhao:2020dvu,Ferreira:2020kvu}. 

At finite (but still not large) temperatures, the uncertainty in the EoS increases. In this case, information about deconfinement can be obtained from supernovae \cite{Fischer:2017lag,Zha:2020gjw} and compact star merger simulations \cite{Most:2018eaw,Bauswein:2018bma,Weih:2019xvw}. Unfortunately, not many finite temperature EoSs allowing for deconfinement to quark matter are available for testing. Additionally, some numerical relativity simulations may run into numerical difficulties when experiencing extremely sharp phase transitions, which could also limit current studies.  

\begin{figure*}
    \centering
    \includegraphics[width=0.5\textwidth]{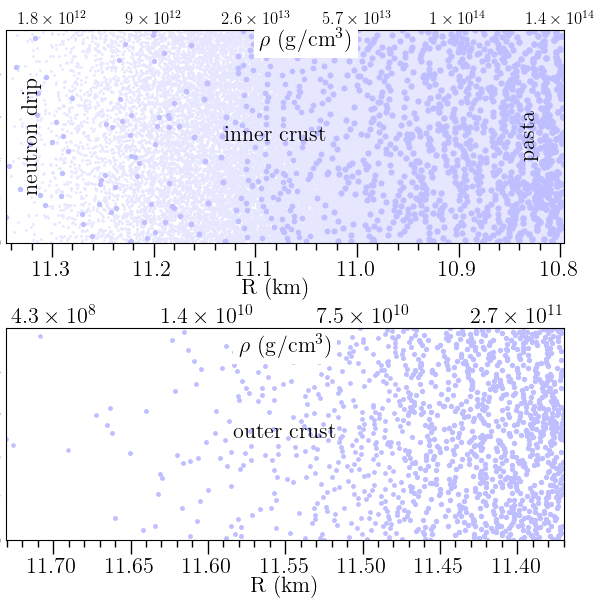}
    \caption{Realistic simulation for the neutron-star crust-core liquid-gas phase transition at zero temperature from \cite{Steiner:2012bq}. The baryon number density and stellar radius are shown. The purple dots represent bulk hadronic matter and the empty white space marks the space occupied by atomic nuclei.}
    \label{crust}
\end{figure*}

Ultimately, in order for EoSs to be able to describe the entire QCD phase diagram and reproduce the smooth crossover predicted by lattice QCD, models must contain both hadronic and deconfined degrees of freedom. At least, this is the prescription that has been used to describe the liquid-gas phase transition expected to take place in the crust-core interface of neutron stars. In the latter, the competition (of the strong force) with the electromagnetic force turns the first-order phase transition into a gradual one and the so called ``pasta phases" appear, as shown in Fig.~3. Comparisons with low-energy heavy-ion collisions are also important to constrain the finite temperature dense regime, but they require modifications to describe zero-net strangeness matter without chemical equilibrium and charge neutrality. As it was shown in  \cite{Aryal:2020ocm,Costa:2020dgc}, these changes in environment conditions can modify the deconfinement position by several tens of MeV for a given temperature. Additionally, there are other considerations when comparing heavy-ion collisions to neutron star mergers that caution against direct comparisons with current theoretical frameworks, as discussed in Sec.\ \ref{sec:dyn}.


\section{Model agnostic approaches to extracting the EoS from gravitational wave data}

Due to the lack of first principle results that can fill the QCD phase diagram in the regime relevant for neutron stars, an alternative approach is to extract a band of possible EoSs from experimental data.  Such an approach attempts to explore the relevant part of the EoS phase space that is causal (the speed of sound is bounded by zero and 1) by fitting observables that depend on the EoS to a set of observations. More specifically, NICER observations of the radiation emitted by hot spots on the surface of rotating neutron stars can be fitted to a pulse profile model \cite{Miller:2019cac,Riley:2019yda,Silva:2020acr}. This model depends on several parameters, including the neutron star compactness (the mass divided by the radius of the neutron star), which in turn depends on the EoS. Therefore, measurements of the compactness place a constraint in the mass-radius plane, which can be converted into a constraint on the EoS. Similarly, LIGO/Virgo observations of the gravitational waves emitted by inspiraling and merging neutron stars can also be used to constrain the EoS \cite{Miller:2019cac,Riley:2019yda,Silva:2020acr}. The LIGO/Virgo data is fitted to a gravitational wave model that depends on several parameters, including the so-called tidal deformabilities, which measure how much a star deforms when in the presence of a certain external perturbation. A measurement of the tidal deformabilities can then be converted into a constrain on the mass-radius plane through the use of certain quasi-universal relations~\cite{Yagi:2013awa,Yagi:2013bca,Yagi:2016bkt,Yagi:2016ejg,Yagi:2015pkc}, which in turn place constraints on the (so far) low and intermediate density portion of the neutron star EoS.

When fitting gravitational wave observations to a model, another approach is to use a parameterization of the EoS. Examples of this parameterization include piecewise polytropes that are patched together at different transition densities~\cite{Read:2008iy}, or a parametric spectral representation \cite{Lindblom:2010bb,Lindblom:2012zi,Lindblom:2013kra}. In both cases, the EoS is assumed to be given by either a known function or a set of differential equations, which depend on a set of parameters. Given a choice of these parameters and a choice of the central density, the Einstein equations can be solved numerically to find the mass, radius and tidal deformability of a neutron star. The waveform model parameter list -- i.e.~the parameters that characterize the waveform model -- is then enhanced to include these EoS parameters, while removing the tidal deformability from the list, which are then varied when fitting the model to the data. From the posterior probability distribution obtained for these EoS parameters, one can then reconstruct an ``allowed'' band in the pressure-density plane for the EoS \cite{Abbott:2018exr,Abbott:2020khf}. 

These parametric approaches, however, suffer from the problem of restricting the EoS to a particular functional form, which may or may not be faithful to the EoS of nature inside neutron star cores. In fact, it has been recently pointed out that piecewise polytropes and the spectral functions both fail to accurately capture bumps, kinks, and jumps in the EoS that may occur due to rapid changes in the degrees of freedom or phase transitions \cite{Tan:2020ics}.  In view of this, a different approach has been recently proposed: a non-parametric model. In this approach, one does not restrict the functional form of the EoS, but rather, one uses Gaussian processes conditioned on nuclear theory models \cite{Essick:2019ldf,Landry:2020vaw} to generate a wide variety of realizations of the EoS. For each of these, the Einstein equations are then solved to compute the mass, radius and tidal deformability, given a choice of central density, which are then used in a waveform model to fit the LIGO/Virgo data. In this way, one circumvents some of the earlier issues mentioned above and provides different confidence regions for the extracted EoS.

One should consider these approaches (the parametric and the non-parametric ones) as attempts to understand the EoS from an ``experimental'' perspective.  They can provide important information about the allowable phase space but cannot provide detailed information about the correct microscopic degrees of freedom.  Thus, such approaches cannot be directly tested against neutron-star cooling data. Temperature evolution is very sensitive to the proton and hyperon content \cite{Beloin:2018fyp,Negreiros:2018cho,Tolos:2016hhl} and, therefore, it is also an important part of EoS modeling at high densities. Additionally, the calculations of transport coefficients \cite{Alford:2019qtm}  also require detailed microscopic information.  This emphasizes that both effective models are needed to understand the microscopic degrees of freedom, but also model-agnostic approaches are needed in order to determine the allowed phase space and help to further constrain microscopic models. 

The use of non-parametric models together with more informed models will become ever more prescient in the coming years, as gravitational wave detectors are upgraded and their sensitivities are increased. The advanced LIGO + (A+) upgrade is expected to conclude by 2025, bringing a plethora of sensitivity improvements to both the Hanford and Livingston LIGO detectors. Of particular relevance are upgrades related to quantum squeezing, which will bring improvements to the high-frequency part of the sensitivity curve. Lowering the noise at high frequencies would allow us to observe the merger of neutron stars, which would in turn allow for more stringent constraints on the EoS and to peek into its low temperature dependence during an out-of-equilibrium process. The extraction of information from the merger phase, however, will require its characterization in terms of numerical relativity models, which must include all of the complexity of the EoS at finite temperature and out-of-equilibrium phenomena. Much work is therefore still needed both in the development of such numerical relativity codes and in the construction of finite-temperature EoSs in the high-density portion of the QCD phase diagram.    


\section{Out of equilibrium effects at large baryon density}\label{sec:dyn}

\subsection{Heavy ion Collisions}

While heavy ion collisions at low beam energies have comparable baryon densities to those reached within neutron stars, we now write a word of caution about direct comparisons between the two fields.  Since the first collisions at RHIC turned on in the early 2000's, numerous discoveries have been made that have dramatically changed our approach to simulating heavy ion collisions.  The field of heavy ion collisions was the first in the world to numerically solve relativistic viscous fluid dynamic equations of motion in complex situations \cite{Romatschke:2007mq} and, in recent years, the necessity of the inclusion of both shear and bulk viscosities \cite{Denicol:2010tr,Noronha-Hostler:2013gga,Noronha-Hostler:2014dqa,Denicol:2014mca,Ryu:2015vwa,Bernhard:2016tnd,Bernhard:2019bmu} (and at large densities BSQ diffusion \cite{Denicol:2012cn,Greif:2017byw,Denicol:2018wdp,Fotakis:2019nbq}) has become clear.  Additionally, since the seminal works nearly a decade ago \cite{Takahashi:2009na,Alver:2010gr}, a deeper understanding has emerged concerning the initial conditions of heavy ion collisions and the role played by quantum fluctuations in the wave function of the colliding nuclei (and potential effects from proton substructure  \cite{Mantysaari:2016ykx,Albacete:2017ajt}). These  radically changed how experimentalists measure flow harmonics and, subsequently, invalidated earlier attempts at describing flow data at low beam energies when theoretical models did not incorporate event-by-event fluctuations.  Finally, our understanding of the influence of critical phenomena has progressed enormously since the first discussions on the search for the QCD critical point \cite{Stephanov:1998dy}.

\begin{figure*}
    \centering
    \includegraphics[width=0.9\textwidth]{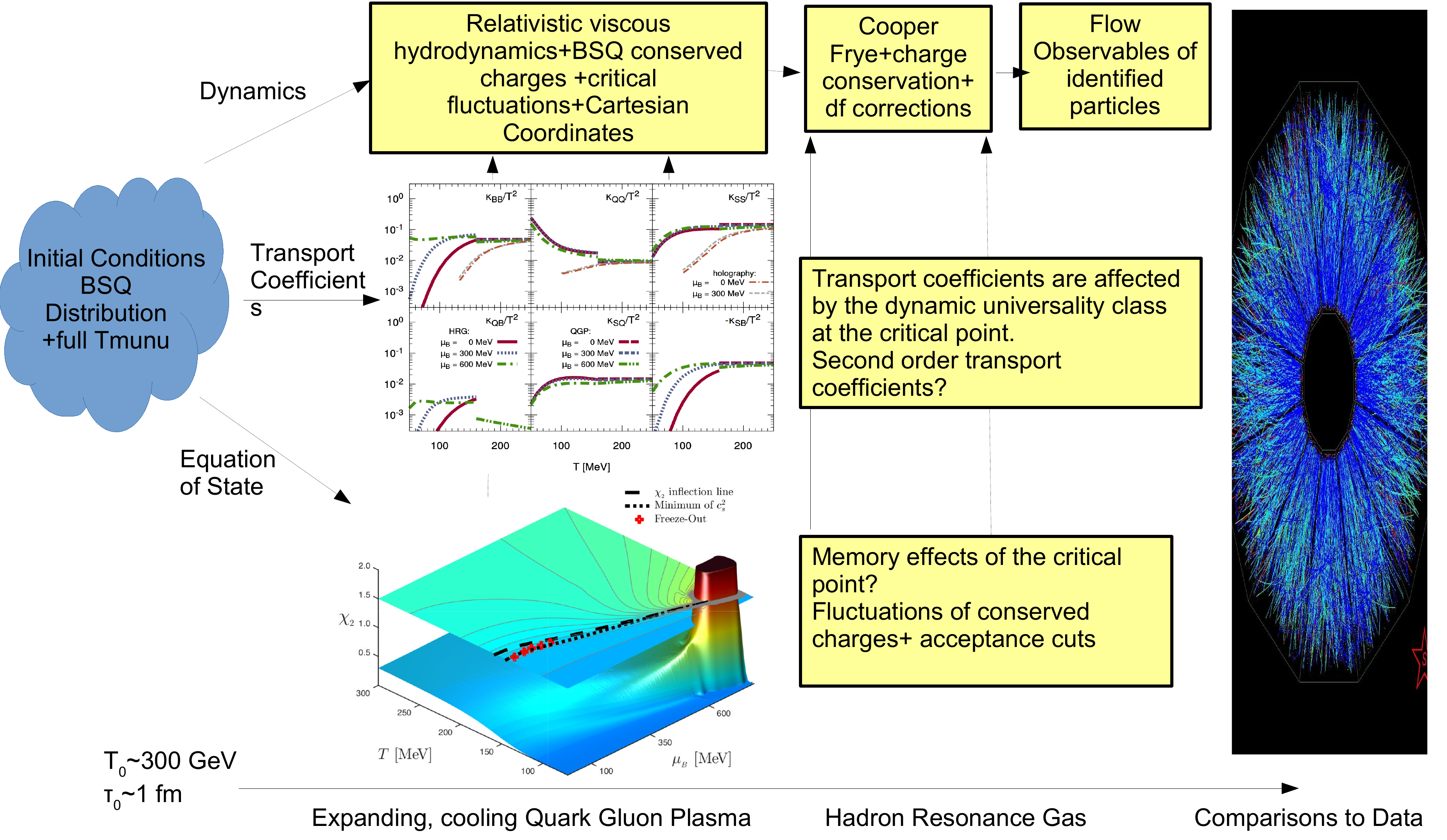}
    \caption{Schematic diagram of the needed changes within heavy ion simulations in order to study the QCD critical point and the first-order phase transition line. The example EoS is taken from \cite{Critelli:2017oub} and the diffusion matrix is taken from \cite{Greif:2017byw,Fotakis:2019nbq}.}
    \label{fig:scheme}
\end{figure*}

Because the heavy-ion field has been primarily focused on high beam energies, a number of upgrades must be implemented in dynamical models before one can attempt to extract information about the EoS from heavy ion data.  An outline of such upgrades is made in Fig.\ \ref{fig:scheme} and is discussed in detail below.  Current models that are applicable at these beam energies only consist of ideal hydrodynamics \cite{Rischke:1995ir,Rischke:1995mt,Aguiar:2000hw,deSouza:2015ena} and, yet, it is known that out-of-equilibrium effects can play a significant role in the search for the QCD critical point \cite{Berdnikov:1999ph,Nahrgang:2011mg,Mukherjee:2015swa,Mukherjee:2016kyu,Nahrgang:2018afz,Dore:2020jye} and the first-order phase transition line \cite{Feng:2018anl}. Thus, the upgrades are a necessary step forward in order to allow for direct theory to experimental data comparisons.  To do this, the initial conditions will need to be adapted to incorporate all 3 conserved charges and be initialized with a full energy-momentum tensor $T^{\mu\nu}$ and $q^{\mu}$'s (out-of-equilibrium effects in the diffusion currents). See \cite{Steinheimer:2008hr,Shen:2017bsr,Martinez:2019rlp,Martinez:2019jbu,Mohs:2019iee,Akamatsu:2018olk} for recent efforts along those lines, although no single initial condition model can accomplish all of that at the time being. 

The effects of critical fluctuations were discussed extensively in the workshop.  One approach is to use Hydro+ \cite{Stephanov:2017ghc}, which provides a generic extension of hydrodynamics by the inclusion of a parametrically slow
mode and fluctuations out of equilibrium.  Other theoretical approaches have been explored as well \cite{Akamatsu:2018vjr,An:2019csj}. The inclusion of such fluctuations in realistic relativistic viscous hydrodynamics models in 2+1 or 3+1 dimensions is a very complex task and steps in this direction can be found in \cite{Young:2014pka,Sakai:2017rfi,Singh:2018dpk,Du:2019obx}.

In fact, event-by-event relativistic viscous hydrodynamic models have been constructed for large beam energies where boost invariance is a reasonable approximation, such that hyperbolic coordinates are used and boost invariant 2+1 hydrodynamics is employed \cite{Romatschke:2017ejr}.  At low beam energies, Cartesian coordinates are more appropriate to describe the fully 3 dimensional evolution of the system.  Unfortunately, in this regime only ideal hydrodynamic codes exist written in Cartesian coordinates, since most viscous hydrodynamics models were written with hyperbolic coordinates \cite{Luzum:2008cw,Schenke:2010nt,Schenke:2010rr,Noronha-Hostler:2013gga,Karpenko:2013wva,Noronha-Hostler:2014dqa,Shen:2014vra}. Thus, this has been identified as an important (and achievable) goal for the future. 

Because in heavy ion collisions baryon number, strangeness, and electric charge conserved currents are relevant, especially at finite densities, hydrodynamics must incorporate all 3 conserved charges and their corresponding diffusion transport coefficients. However, since individual quarks can carry two or three charges (for example, a strange quark carries electric charge, strangeness, and baryon number), there are cross-correlations between the diffusion transport coefficients, which actually become a diffusion matrix \cite{Greif:2017byw}.  A number of groups are already addressing this issue and we will likely have working BSQ hydrodynamic codes available in the years to come. However, this also requires special attention at the point of freeze-out such that both BSQ charges are conserved \cite{Oliinychenko:2019zfk} and the appropriate out-of-equilibrium corrections are taken into account (this is known in the field as $\delta f$ corrections).  

Finally, at large densities one requires an accurate description of the hadron gas phase, for which one must use transport codes that incorporate the known hadrons and their interactions.  SMASH \cite{Weil:2016zrk} has become the state-of-the-art in the field for hadronic transport, and it also incorporates mean field potential effects. One should note that there are other hadronic transport codes on the market and each of their underlying assumptions produces different results \cite{Zhang:2017esm}. Once a hadronic model is established, one can use it to calculate transport coefficients as well \cite{Rose:2017bjz}. 

In principle, other considerations would also go into dynamical models such as spin \cite{Weickgenannt:2019dks,Florkowski:2019voj,Weickgenannt:2020aaf,Bhadury:2020puc,Montenegro:2020paq} and magnetic field effects \cite{Skokov:2009qp,Fukushima:2008xe,Jiang:2016wve,Finazzo:2016mhm,Shi:2017cpu,Sheng:2018jwf,Oliva:2019kin,Denicol:2019iyh}. While there has been very intriguing experimental data indicating that the quark-gluon plasma may be the most vortical fluid known to humanity \cite{STAR:2017ckg}, further studies are needed in this regard to draw strong conclusions at this time. Finally, once a full viscous hydrodynamic model that includes transport is established, there will be a number of interesting observables to compute, such as directed flow, which is expected to be sensitive to the EoS \cite{Nara:2016phs}

\subsection{``Heavy-ion Data" to constrain Neutron Stars }

One topic that was brought up at the workshop was the use of ``heavy-ion data", extracted from the insightful phenomenological study done in \cite{Danielewicz:2002pu}, as a constraint on the neutron-star EoS. We use quotation marks because this study was not performed by an experimental collaboration. Rather, the results in \cite{Danielewicz:2002pu} constitute a theoretical framework to extract EoS constraints within specific model assumptions and, therefore, cannot be considered bona fide experimental data. In fact, there are a number of  assumptions (reasonable at that time) that went into this extraction of constraints that are not universally agreed upon by the entire heavy-ion community after two decades. Currently, the standard way to extract information from experimental data within heavy ions is to use event-by-event relativistic viscous hydrodynamics and make direct comparisons to experimental data using, for example, a Bayesian analysis (e.g. \cite{Bernhard:2019bmu}).  Such a framework does not yet exist for data at low beam energies due to the issues outlined above and laid out in Fig.\ \ref{fig:scheme}.  While initial attempts have been made for comparing ideal hydrodynamics \cite{Spieles:2020zaa} to experimental data at low beam energies, one must incorporate in this case transport coefficients, as there are many indications that the influence of out-of-equilibrium effects only grows with lower beam energies. Thus, we  emphasize that the problem is not   Ref.\ \cite{Danielewicz:2002pu} but rather the misuse of their results as a de facto experimental heavy ion constraint that must be used to eliminate certain physical scenarios concerning the high density regime of the EoS in neutron stars. 

Instead, what has usually been done is to either compare hadronic  models (transport) to experimental data and attempt to extract the EoS from such models (this was done in \cite{Danielewicz:2002pu,Hillmann:2019wlt}) or to compare ideal hydrodynamics to data with different EoS assumptions, as was done in \cite{Spieles:2020zaa}.  At the moment, neither are a perfect fit to experimental data and fail to capture a number of experimental observables.  However, we point out that the EoS that does fit best the results from hydrodynamics studies is consistent with the presence of quark degrees of freedom at the low heavy ion beam energies.  Additionally, we stress that HADES is an ongoing experiment running at $E_{lab}=1-2$AGeV, right in the middle of the beam energies from \cite{Danielewicz:2002pu}, and it has found evidence that the temperatures reached in their reactions are much higher than initially anticipated, which makes the previous indication of reaching quark degrees of freedom much more likely \cite{Adamczewski-Musch:2019byl}.  Furthermore, hints from STAR and HADES find long range correlations in multi-particle cumulants of proton number, which may be providing hints of a real phase transition  \cite{Adam:2020unf,Adamczewski-Musch:2020slf}. 

Thus, our current understanding of heavy ion collisions, which has evolved immensely in the last two decades, strongly indicates that the use of the ``heavy-ion data" extracted from the influential Ref.\ \cite{Danielewicz:2002pu} as a constraint for the neutron star EoS can be misleading, as there are many known discrepancies in the field that will likely be attributed to a variety of needed upgrades to dynamical models, as outlined in Fig.\ \ref{fig:scheme}. Additionally, experimental data from HADES indicate that even at very low beam energies there are signs of deconfined quarks and gluons. However, many more theoretical studies are still required for that to be a solid statement. In summary, one should not find this to be a disappointing conclusion but rather we emphasize that more cross-talk is needed between communities in order to understand the state-of-the-art in the respective fields, so that better and more meaningful comparisons can be made. 

\subsection{Neutron Star Mergers}

It was pointed out in the workshop that multi-messenger observations of neutron star mergers \cite{TheLIGOScientific:2017qsa,Monitor:2017mdv,GBM:2017lvd} give not only key information on the thermodynamical behavior of dense matter, but they also provide fundamental insight into the complex out of equilibrium processes that take place on millisecond timescales. In fact, transport properties are a better discriminator of different phases than the EoS \cite{Alford:2007xm}. For neutron star mergers, the important dissipation mechanisms are those whose equilibration times are $\leq 20$ ms, as they can affect the post-merger gravitational wave signal. 

For many years it was assumed that the evolution of the  hot and ultradense matter formed after the merger could be reasonably described as an ideal fluid (dynamically coupled to Einstein’s equations) since the time scales for viscous transport to set in were previously estimated \cite{Bildsten:1992my} to be outside
the millisecond range. These estimates were recently revisited in \cite{Alford:2017rxf} using state-of-the-art merger simulations and it was concluded that damping of high-amplitude oscillations due to bulk viscosity is likely to be relevant if direct Urca processes remain suppressed. Thus, bulk viscosity effects should be investigated in neutron star merger simulations \cite{Alford:2017rxf}. Further studies about the bulk viscosity in the context of mergers were performed in  \cite{Alford:2019qtm,Alford:2019kdw,Alford:2020lla,Alford:2020pld}. Ref.\ \cite{Alford:2017rxf} also concluded that neutrino-driven thermal transport and shear dissipation were unlikely to affect the post-merger gravitational wave
signal unless turbulent motion occurs. 

In this context, as pointed out in  \cite{Duez:2004nf}, viscosity and magnetic fields drive differentially rotating stars toward uniform rotation, which has important consequences. For instance, a differentially rotating hypermassive remnant can momentarily support a mass greater than it would be possible for a uniformly rotating star, which would imply in the observation of a delayed collapse to a black hole and a delayed burst of gravitational radiation. However, the magnetic field in the differentially rotating remnant may be amplified through magnetorotational instabilities \cite{Balbus:1998ja}, which can generate magnetohydrodynamic (MHD) turbulence whose description requires extremely high resolution simulations \cite{Kiuchi:2014hja}. General-relativistic shear viscous hydrodynamics becomes an interesting phenomenological alternative for studying how angular momentum transport occurs in such a system \cite{Duez:2004nf,Shibata:2017xht}, with the effective viscosity being induced by local MHD turbulent processes.

As stressed in \cite{Alford:2017rxf}, the effects of bulk, shear, and thermal conductivity  have not yet been included in merger simulations because that requires a formulation of general-relativistic viscous fluid dynamics that is compatible with causality in the strong nonlinear regime probed by the mergers. Fortunately, a fully consistent physical and mathematical description of viscous fluids dynamically coupled to Einstein's equations is now possible as recently proven in \cite{Bemfica:2020zjp}. The latter employed the basic ideas behind the new formalism originally proposed in \cite{Bemfica:2017wps} to obtain a causal and stable first-order generalization of the relativistic Navier-Stokes equations for the case of a conformal fluid. This new approach to relativistic viscous hydrodynamics was later further developed and generalized in  \cite{Kovtun:2019hdm,Bemfica:2019knx,Hoult:2020eho} to include the non-conformal regime and also finite baryon density effects.

Among many other results, it was also discussed in the workshop how prompt black holes formed in neutron star mergers can source bright electromagnetic counterparts for high-mass-ratio binaries, which can be constrained from multimessenger observations \cite{Bernuzzi:2020txg}. Nevertheless, it was thoroughly emphasized in the meeting that, while the inspiral~\cite{Dietrich:2018uni} and early postmerger phases (in the case of a black hole remnant) are now better understood, there is still a vast parameter space to explore. In fact, no current simulation of neutron star mergers include all the relevant physics, as it becomes increasingly complex on longer timescales in the postmerger and higher resolution, and more sophisticated simulations, are needed.

\section{Outlook}

After three days of a very intense (virtual) workshop, it became clear that there is a very large number of  challenges and fundamental questions that must be addressed when it comes to determining the properties of ultradense matter in heavy ion collisions and also in neutron stars. The following questions were singled out and discussed during the meeting:

\begin{itemize}
    
\item What is the nature of matter at high baryon densities/chemical potentials? How can we measure QCD critical phenomena? Are there new exotic phases in neutron stars? Can we probe these with gravitational waves from neutron star inspirals and mergers?

\item What are the prospects of making a 10\% measurement of the tidal deformability of a neutron star with advanced LIGO? What are the challenges posed by detector upgrades and waveform systematics? And what nuclear physics information, other than pressure versus energy density (EoS), are nuclear physicists interested in knowing that one could extract from the tidal deformabilities or from the a merger phase?

\item How far can we push our understanding of hadronic matter and its interactions at high densities? How does one model and realistically constrain the EoS at finite temperatures and densities? What is the interplay between the EoS probed in heavy ions and that probed in neutron stars and their mergers? For instance, if mergers reveal that there is a first order phase transition at finite temperatures, this proves that the QCD critical point exists. 

\item What is the best approach to extract an EoS from neutron star inspirals and NICER data? What do we miss by just focusing on the EoS? For instance, for cooling and transport one needs to know the correct phase of matter, including degrees of freedom and interactions.

\item How can hydrodynamics and transport be used to determine the correct degrees of freedom in low energy heavy-ion collisions and neutron star mergers? What upgrades are needed for current codes to be able to extract the EoS from low beam energies? What role do out-of-equilibrium effects play at larger densities and at a critical point/phase transition? What role does isospin/strangeness play in their respective dynamics? What signals from nuclear experiments are most promising to extract the EoS and information about phase transitions? 

\item What are the challenges that must be faced when performing realistic simulations of ultradense matter in heavy ion collisions and in neutron star mergers? What is the synergy between them? What are the tools needed to systematically investigate out-of-equilibrium viscous effects under extreme temperatures, densities, and gravitational/electromagnetic fields?

\item What are the consequences beyond QCD i.e. if we understand the QCD EoS across the entire phase diagram? Can one then make quantitative statements about dark matter? 

\end{itemize}

Clearly, these questions are far too complex to be tackled by one community alone – only truly interdisciplinary collaboration will be the herald of progress. In fact, the authors hope new insights and collaborations will emerge from this and future workshops to help solve the challenging questions mentioned above.

\section*{Acknowledgements} We thank all the speakers, panelists, and participants of the virtual workshop ``From heavy-ion collisions to neutron stars" for illuminating discussions and the Illinois Center for Advanced Studies of the Universe (ICASU) for support. V.D. acknowledges support from the National Science Foundation under grant PHY-1748621. J.N. is partially supported by the U.S. Department of Energy, Office of Science, Office for Nuclear Physics under Award No. DE-SC0021301. J.N.H. acknowledges support from the US-DOE Nuclear Science Grant No. DE-SC0019175 and the Alfred P Sloan Foundation. N.Y. thanks the Illinois Center for Advanced Studies of the Universe from the Department of Physics at the University of Illinois at Urbana-Champaign for support. This  material  is  based  upon  work  supported  by  the National  Science  Foundation  under  
grant No. PHY-1654219 and by the U.S. Department of  Energy,  Office  of  
Science,  Office  of  Nuclear  Physics, within the framework of the Beam Energy Scan Theory
(BEST) Topical Collaboration.  

\bibliography{mybib.bib}

\end{document}